\documentclass{sbrt}
\usepackage[english,brazil]{babel}
\usepackage[utf8]{inputenc}

\usepackage{graphics}
\usepackage{graphicx}
\usepackage{graphics}

\begin{document}

\title{The Behavior of Internet Traffic for Internet Services during COVID-19 Pandemic Scenario}

\author{Carlos Alexandre Gouvea da Silva, Allan Christian Krainski Ferrari, Cristiano Osinski and Douglas Antonio Firmino Pelacini
\thanks{Carlos Silva, Computer Science Department, State University of Parana Midwest, Unicentro, Guarapuava, Brazil, e-mail: carlos.gouvea@ieee.org; Allan Ferrari, Centro Universit\'{a}rio Unicuritiba, Curitiba, Brazil, e-mail: allan.ferrari@unicuritiba.com; Cristiano Osinski and Douglas Pelacini, Department of Electrical Engineering, Electrical Engineering Graduate Program, Federal University of Paran\'{a}, Curitiba, Brazil, e-mail: cristiano.osinski@ufpr.br and douglaspelacini@ufpr.br.}%
}

\maketitle

\markboth{XXXIX SIMP\'{O}SIO BRASILEIRO DE TELECOMUNICA\c{C}\~{O}ES E PROCESSAMENTO DE SINAIS - SBrT 2021, 26--29 DE SETEMBRO DE 2021, FORTALEZA, CE}{}


\begin{abstract}
Since the end of 2019, the SARS-CoV-2 virus known as COVID-19 has spread rapidly around the world, forcing many governments to impose restrictive blocking or lockdown to combat the pandemic. With locomotion restriction of people in almost of countries of the world, workers and students needed to keep their activities at home. As a result, people's behavior, habits, and the way they started using the Internet changed significantly. Like professionals of offices, the younger played an important role in this behavior, especially in the type of resources used by them. As result, the characterization and traffic of communication networks were affected in some way. In this perspective article, we join from many available studies about the COVID-19 effect at networks and investigate the effects on the Internet traffic of using services such as video streaming, video conferencing, and gaming during 2020's months of the pandemic.
\end{abstract}

\section{Introduction}

Since the 2000s, the world has been affected by viruses that badly attack mainly the respiratory system of people, which can lead to the death of many of them. 
Among this virus, in 2003 the Severe Acute Respiratory Syndrome (SARS), in 2009 H1N1 and then in 2012 the Middle East Respiratory Syndrome (MERS) were the main viral agents that have plagued the world since then \cite{Mari:2020}. 
Since 2000s years The new 2019 coronavirus (COVID-19) or severe acute respiratory syndrome coronavirus 2 (SARS-CoV-2) has become a global concern, and spread rapidly since its origin in the city of Wuhan, Hubei province, China \cite{Wang:2020}.

The first cases of the new COVID-19 are reported from December 2019, in which adults in Wuhan started performing at local hospitals with severe pneumonia with unknown causes. 
Many of these cases were similar to the exposure of these people in the seafood market in the city, which also reports the sale of live animals \cite{Singhal:2020}.

After an investigation of the origin of the new virus by surveillance systems (installed after the SARS outbreak 2002-2003) \cite{Singhal:2020}, on December 31, 2019, China notified the outbreak to the World Health Organization (WHO) and on January 1, 2020, closed the seafood market of the Huanan. 
Then, WHO declared the outbreak a public health emergency of international concern \cite{Du:2020}. 
One of the most challenges was to keep the spread of COVID-19 in China under control in order to prevent it from spreading to other regions and countries in other continents \cite{Boldog:2020}. 
Still, this scenario presents the crucial importance of the main public health interventions in the control of COVID-19 epidemics in China or in other countries. On 11 February 2020, WHO officially declared the disease caused by the COVID-19, and on March 11, 2020, the WHO declared the new virus outbreak a world pandemic \cite{Cucinotta:2020}.

As precautionary action to the health of workers, old people, risk group people, children, students, and also with the objective of reducing the proliferation of the virus in environments with agglomeration, the governments of each country took immediate actions. In most cases, governments have decided to close non-essential services, offices, schools, and universities as an action to mitigate the impacts of COVID-19 on the health systems of each country. 

In this way, office tasks (Bloom, Davis and Zhestkova, 2020) and face-to-face classes would be held remotely \cite{Gonzalez:2020,Khan:2020}. 
Several digital distance learning platforms allow interaction between teachers and students and, in some cases, national public television programs or social media platforms are being used for education during the pandemic period \cite{Gonzalez:2020}.

Remote work has been used by companies since the 1970s, in which this type of work represented only 5\% in 2019, however with the new COVID-19 companies had to change this new model of work quickly. In this new model of remote work, workers can share files, in addition to attending meetings with audio and video \cite{Leonardi:2020}. 
Still, the new COVID-19 brings new challenges and opportunities for workers with disabilities to telework at home \cite{Schur:2020}.

With the increase in the number of users in isolation due to the need for lockdown, there was an increase in the number of users of technological services such as video, voice, streaming and educational resources. In addition, technology is being used to help relieve the stress and anxiety caused by the pandemic using online games for entertainment \cite{KirAly:2020}. 
Many of these technologies requires a minimum of bandwidth and throughput in order to keep minimum of quality on the services at Internet, resulting a considerable requirements of the Internet traffic or characterization in core properties, such as suggested by \cite{Stepanenko:2002}. 
However, studies also show parameters such as latency of Internet was affected caused by the increased amount of human activities that are carried out on-line for work or study \cite{Candela:2020}.

This paper present an overview about how the new COVID-19 affect the Internet Traffic in different countries and services available at Internet.
This study is based in a set of published papers in 2020 where Internet traffic was analyzed due COVID-19 pandemic.   
We provide a discussion about the impact of lockdown effect at Internet traffic when hundred of millions of citizens were forced to stay at home for remote working, entertainment, commerce, and education. 

Including this introductory section, the section 2 presents a short view about Internet behavior before COVID-19, and then the section 3 present the impact of COVID-19 pandemic at Internet traffic.

\section{Internet Traffic before new coronavirus}

Still in March 2020, before the WHO announcement that COVID-19 had become a global pandemic, Cisco Telecommunication company made available in its annual report the data for global forecast/analysis that assesses digital transformation across various business segments \cite{Cisco:2020a}. 
In this report (2018-2023) it is described that changes in the pattern of user devices, the types of connections and also the growth in ownership of various devices affect Internet traffic patterns. Video devices can have a traffic multiplier effect, where videos in UHD (ultra high definition) or 4K resolution can have up to twice the bit rate of HD (high definition) videos and nine times more than standard definition (SD) video \cite{Cisco:2020a}.

Improvements in broadband speed result in increased data consumption and use of high-bandwidth content and applications, such as video and streaming content. The global average broadband speed is expected to grow and more than double from 2018 to 2023, from 45.9 Mbps to 110.4 Mbps \cite{Cisco:2020a}.

According \cite{Cisco:2020b} in 2022 Internet video will represent 82\% of all business Internet Traffic, where VR/AR traffic will increase twelvefold, and Internet video surveillance traffic will increase sevenfold. Augmented Reality (AR) and Virtual Reality (VR) are becoming common place and many technological companies are investing a lot in this kind of content for end users, i.e., Facebook, Google, Microsoft, among other \cite{Elmqaddem:2019}. 
Many countries have users who use transmission rates above 125 Mbps, paving the way for future video demands. It is predicted that the applications that will demand the most from the network for internet content traffic will be higher for UHD VR, HD VR, 8K wall TV, UHD IP video and cloud gaming. For the first case of UHD VR applications, the minimum network requirements are expected to be a rate of 500 Mbps for service operation \cite{Cisco:2020a}.

It is possible to note that the traffic forecasts on the internet network are based on studies of historical data series and also on estimates generated by the companies responsible for the development of services that will demand these network resources. Still, after the need for lockdown imposed by several countries to prevent the spread of COVID-19, many workers and students started to use technological resources for remote work and study.

\section{Internet Traffic in Times of COVID-19}

The rapid global spread of COVID-19 has increased the volume of data generated from various sources, mainly services based in video and voice content, and cloud computing that allow workers and students to efficiently accomplish their tasks and activities. Cloud computing is formed by platform, hardware, software, and infrastructure where the service is accomplish for users. 

%

Cloud computing applications has increased the volume, velocity, and/or variety in data being generated around the world from several services and application available at Internet \cite{Alashhab:2020}. 
As more people, whether working adults or students, have to work or learn remotely in response to the spread of the COVID-19 pandemic, there is a high demand for popular cloud computing services. For education and business section, it was suggested services like DingTalk, Google Meet, Microsoft Teams, and Zoom for live-video communications platforms \cite{Alashhab:2020}. 
Netflix, Disney Play, HBO Max and Prime Video are the most popular entertainment applications for entertainment in lockdown situation. During the period of the pandemic because of overloads at video services, Netflix needed cutting down the quality of its streaming.

\begin{figure}[!ht]
	\centering 
	\includegraphics[clip,width=8cm]{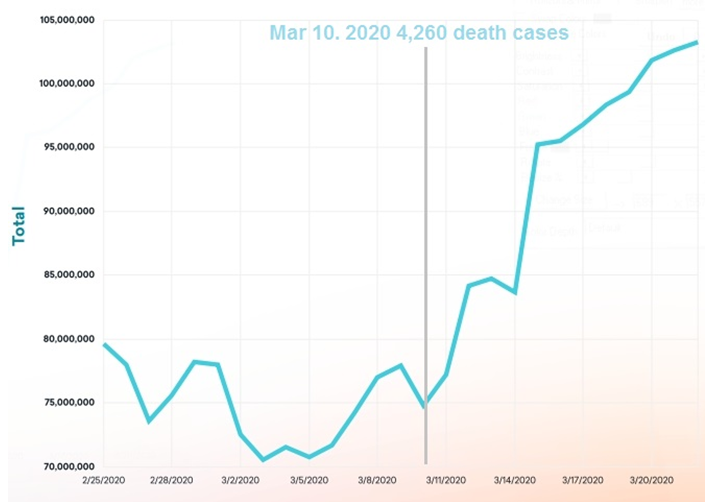} 
	\caption{International Entertainment \& Streaming Sessions: Throughout the Beginning of COVID-19 \cite{Apptopia:2020}.}
	\label{fig1}
\end{figure}

Figure \ref{fig1} presents the growth in the number of streaming section during beginning of COVID-19 in March 10, 2020 where the number of death cases in all the world was only 4,260. According Apptopia report ``\textit{Already, the global pandemic is transforming both streaming behavior and media. In March, we saw a 30.7\% increase in streaming sessions. Along with a week-over-week increase, the industry began to evolve: Pixar announced it would release Onward, an animated movie currently in theaters, to digital marketplaces on March 20 and to Disney+ on April 3. Similarly, on March 22, Amazon launched Prime Video Cinema, a service that allows consumers to rent or buy recently released movies}'' \cite{Apptopia:2020}. 
Apptopia analysis spans August 2019 to March 2020 and looks at 35 streaming services, including YouTube, Amazon Prime Video, HBO GO, HBO NOW, DC Universe, Disney+, ESPN, among others.



The coronavirus pandemic has a major impact on the Internet, because the unexpected demands for the various services used during the lockdown have caused some problems for Internet service providers (ISPs). In some European countries the use of the Internet increased by 50\% and in the United Kingdom this increase was 30\% \cite{Alashhab:2020}. 
The changes in service standards of several companies that broadcast video and entertainment content were aimed at reducing the impact on ISPs. Video conference services, such as Zoom and WebEx, have also been changed in order to reduce the impact on ISPs \cite{Alashhab:2020}. 

\begin{figure}[!ht]
	\centering 
	\includegraphics[clip,width=8cm]{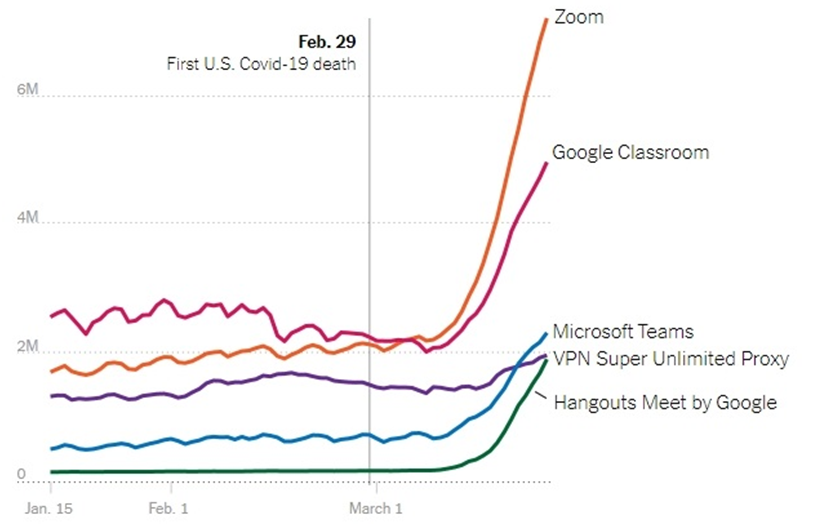} 
	\caption{Daily app sessions for popular remote work apps. Source font: \cite{NYT:2020}.}
	\label{fig2}
\end{figure}

Figure \ref{fig2} presents the daily app sessions for popular remote work apps in USA, where it is possible to see a growth of app section for the main remote work apps, such as Zoom, Microsoft Teams and Hangout Meet for video conference or meeting, and Google classroom for students.


In \cite{Favale:2020} it is provided an analyze of the impact of the lockdown enforcement on the Politecnico di Torino campus network after the lockdown enforcement on February, 25 of 2020 in Italy. The authors focus at the campus traffic, focusing on collaboration and remote working platforms usage, remote teaching adoption, and look for changes in unsolicited/malicious traffic at University \cite{Favale:2020}. 
During lockdown, a 10 times decrease in incoming traffic during the lockdown due restriction of students and professors within University. On the other hand, outgoing traffic grows instead of 2.5 times, driven by more than 600 daily online classes with around 16,000 students per day that follow classes \cite{Favale:2020}. 
The same study presents a comparative Internet traffic between three Italian Universities, where in all cases the incoming traffic before was higher than after of lockdown. Outgoing Internet traffic changed in Politecnico di Torino campus, where an in-house online teaching service has been deployed. This system caused a massive inversion on traffic patterns, where there were a significant growth in upload traffic due to online teaching services, i.e., text files, live classes and record videos \cite{Favale:2020}. 
At same University, \cite{Tropea} show that during weekdays within COVID-19 pandemic, the total bit rate exceeds 1 Gbit/s, with live streaming of classes responsible for a bit more than one-third of the traffic. During the weekend, when no lecture is scheduled, we still observe large traffic (up to 750 Mbit/s) due to students downloading on-demand lectures and teaching material \cite{Tropea}.

While workers and students switched to working and study from home and maintaining social distancing measures, their internet consumption behavior changed as well, especially for specific sector of industry. 
According \cite{Pantelimon:2020}, commercial areas such as e-Commerce was positively affected by COVID-19, where there were a growth of Internet traffic at e-Commerce, Telecom and Retail tech sector, 135\%, 129\% and 76\% respectively. 
These information is based April of 2020 compared to January in Romania and Germany. On the other side, the sector with most decrease of traffic was tourism sector, where presented -73\% of Internet traffic \cite{Pantelimon:2020}.

Internet traffic on ISPs was analyzed in USA by \cite{Liu:2020}, where also was presented data information about growth of traffic in many countries of Europe, such as Italy and UK. 
In the Comcast network ISP the authors find that downstream peak traffic volume increased 13-20\%, while upstream peak traffic volume increases by more than 30\%. Also, they related this significant changes in the magnitude of traffic during the lockdown due COVID-19 pandemic. The same study present changes in Round Trip Time (RTT) at non-satellite ISPs in the Federal Communications Commission Measuring Broadband America program (FCC MBA), where there were a growth at first days of March, 2020 \cite{Liu:2020}. 
Based in video services content in USA: Zoom and WebEx increased advertised IP address space by about 4 times and 2.5 times respectively, roughly corresponding to the 2-3 times increase in video conferencing traffic \cite{Liu:2020}.

In \cite{Bottger:2020} it is presented a overview about the changes Internet behavior during COVID-19 pandemic at Facebook's Edge Network. 
According authors, Internet traffic at Facebook network increased in all continent analyzed in March when compared to previously months. They can see a steady growth was traffic until the second half of March 2020. This internet traffic growth was expected as it reflects the organic growth of the underlying platform and social network during COVID-19 pandemic. Still, the growth of traffic was different between continents, where it is possible to correlated each period based in major outbreaks and countermeasures taken by governments in each regions, e.g., North America and Europe \cite{Bottger:2020}. 
When looking at regions such as Europe (Italy, France, Spain, UK, Switzerland and Belgium), India and the USA, there was a considerable growth in internet traffic caused by the use of live stream services. In Europe this growth was more than 240\%, in India approximately 350\% and in the USA just over 300\%. In addition, services such as messaging also saw traffic of approximately 90\%, 60\% and 50\%, respectively for Europe, India and the USA. In all of these regions there was a small drop in the percentage of photo sharing services \cite{Bottger:2020}.

A data traffic evolution study from 23rd of February until the 10th of May in UK for mobile Internet traffic is carried out by \cite{Lutu:2020}. 
Author observed in local residential ISPs that after lockdown caused by COVID-19 people relied less on the cellular network for data connectivity (e.g., using home Wi-Fi connectivity instead), thus contributing to the surge of traffic reported by residential ISPs \cite{Lutu:2020}. 
Wi-Fi is a connection network infrastructure for wireless local area network with high velocity, low cost, most flexibly and easily maintenance \cite{Silva:2019}. 
For \cite{Lutu:2020} it was noticed overall modest changes for variation of the total uplink data volume per cell in the UK, with, less than 5\% decrease in the median daily volumes in week 18. 
This suggests that applications making intensive use of the downlink (e.g., video streaming) suffered a significantly higher traffic reduction than applications with symmetric uplink/downlink usage (e.g., audio/video conferences, or voice traffic) \cite{Lutu:2020}.

In \cite{Feldmann:2020} it is described the lockdown effect in one ISP from Europe, three IXPs from Europe and USA, and one metropolitan Spanish educational network during COVID-19 pandemic. Internet Exchange Point (IXP) is a network interconnection between Internet providers and content delivery networks and their networks. Relative traffic volume changes follow user changing habits causing "moderate" increases of 15-20\% at the height of the lockdown for the ISP/IXPs, but decreases up to 55\% at the educational network. Even after the lockdown is scaled back, some of these trends remain: 20\% at the IXP from Central Europe but only 6\% at the ISP \cite{Feldmann:2020}. 
The same study shows most traffic increases happen during non-traditional peak hours, and daily traffic patterns are moving to weekend-like patterns. This behavior is related to video conferences, video on demand (VoD) content, gaming traffic, and video streaming.

\section{Conclusion}

In 2020 the world was affected by a new coronavirus, as known COVID-19, where billion of citizens must to keep at homes in order to avoid virus was proliferating. Since the beginning of world lockdown workers and students has been stayed at home where they carried out their activities remotely using technological services at Internet. This brief paper has explored how Internet traffic demands changed as a result of the abrupt daily patterns caused by the COVID-19 lockdown, how these changing traffic patterns affected the performance of ISPs in many countries around the world. In almost of cases these changes was caused by specific services where requires a large band for a minimum operation of quality of services. Services such as video conferences, video meeting, video streaming, gaming, educational and cloud services had a significantly impact at Internet traffic.


\end{document}